\documentclass[a4paper]{jpconf}
\usepackage{graphicx}
\usepackage{comment}
\usepackage{amsmath}

\begin{document}
\title{Q-triplet characterization of atmospheric time series at Antofagasta: A missing values problem}
\author{Hishan Farf\'an-Bachiloglu$^{1^\ast}$, Francisco A. Calder\'on$^{2}$, Dar\'io G. P\'erez$^{1}$}
\address{$^{1}$Instituto de Física, Pontificia Universidad Católica de Valparaíso (PUCV), Valparaíso 23-40025, Chile\\
$^{2}$Departamento de Física, Universidad Católica del Norte, Av. Angamos 0610, Antofagasta, Chile
}

\ead{hishan.farfan@pucv.cl}

\begin{abstract} 
Located in northern Chile (23.7°S, 70.4°W), Antofagasta has an exceptionally arid and stable climate characterized by minimal precipitation and consistent weather patterns. Nevertheless, despite these climate conditions being meaningful for several research and practical applications, our understanding of weather dynamics remains limited. The available meteorological data from 1969 to 2016 is analogical, which presents a significant challenge to analyze because these records are riddled with missing values, some measurements were taken at irregular measuring intervals, making it an interesting puzzle to grasp the Antofagasta's climate scenario.
To overcome this issue, we present a comprehensive statistical analysis of atmospheric temperature, pressure, and humidity time series. Our analytical approach involves the q-triplet calculation method, serving as a powerful tool to identify distinctive behavior within systems under non-equilibrium states.
Our results suggest that, in general, the q-triplet values satisfy the condition $q_{\mbox{\tiny sens}}<1<q_{\mbox{\tiny stat}}<q_{\mbox{\tiny rel}}$, a pattern that has been observed in previous studies.
\end{abstract}

\section{Introduction}
The complexities of dynamic systems, particularly those diverging from thermodynamic equilibrium, pose significant challenges, but also opportunities for scientific exploration. The $q$-triplet method, derived from Tsallis's non-extensive statistical mechanics, has emerged as a robust and novel tool for unraveling the intricate behavior of such systems~\cite{Burlaga2005}. This study delves into atmospheric time series, focusing specifically on daily measurements of atmospheric pressure, air temperature, and relative humidity at 08:00, 14:00, and 20:00 hours from 1969 to 2016. The choice of these data is driven by their essential role in studying the atmospheric dynamics of Antofagasta city, sourced from the only meteorological station embedded in the city. These data are pivotal for examining the city's climate, characterized by its stability and aridity~\cite{Rutllant2003}. Notably, the $q$-triplet method allow us to study the system despite the significant gaps in the data, adding a unique dimension to this research.

Analyzing these data using the $q$-triplet method is (not only) a statement to the method's versatility across various scientific disciplines but also responds to the growing need for advanced analytical tools for studying systems outside equilibrium. Previous research, such as that conducted by Ferri et al. (2010) on the stratospheric ozone layer ~\cite{Ferri2010} and, also by Stosic and colleagues in hydrological dynamics~\cite{Stosic2018}, had demonstrated the efficacy of the $q$-triplet method in extracting meaningful insights from complex datasets. Our study aims to extend this body of work by applying the $q$-triplet method to these atmospheric series, thereby contributing to a deeper understanding of the stability of this climate scenario.

By expanding our analysis to include a series of diverse natures, we endeavor to provide a comprehensive perspective that integrates the statistical prowess of the $q$-triplet with the inherent complexities of atmospheric data. This approach enhances our understanding of micrometeorology and sets the stage for future explorations in similar complex systems.

\section{Methodology}

In the application of Singular Spectrum Analysis (SSA)~\cite{Golyandina2013} to our time series data, we commence by embedding the original time series $\{X_t\}_{t=1}^N$ into a trajectory matrix $\mathbf{X}$. This matrix was constructed by arranging lagged length vectors $L$ in columns, thus forming a Hankel matrix of dimensions $L \times K$, where $K = N - L + 1$. Subsequently, Singular Value Decomposition (SVD) was performed on $\mathbf{X}$, decomposing it into $\mathbf{X} = \mathbf{U} \mathbf{\Sigma} \mathbf{V}^\top$, where $\mathbf{U}$ and $\mathbf{V}$ are orthogonal matrices, and $\mathbf{\Sigma}$ is a diagonal matrix containing singular values. The principal components is then obtained by projecting the original data onto the eigenvectors in $\mathbf{U}$. We isolated the signal representing the underlying trend by reconstructing the series using selected principal components, thereby obtaining the residuals.

The q-triplet method, rooted in Tsallis's non-extensive statistical mechanics, focuses on the analysis of three fundamental parameters: $q_{\mbox{\scriptsize stat}}$, $q_{\mbox{\scriptsize rel}}$, and $q_{\mbox{\scriptsize sens}}$. These reflect critical aspects of the behavior of complex systems: $q_{\mbox{\scriptsize stat}}$ is related to the probability distribution, $q_{\mbox{\scriptsize rel}}$ to the rate of relaxation towards a steady state, and $q_{\mbox{\scriptsize sens}}$ to the sensitivity to initial conditions.

To calculate the $q_{\mbox{\scriptsize stat}}$ using the method described by Ferri et al. (2010)\cite{Ferri2010} but adapted to the specifics of our study, we begin with the deseasonalization of the time series to obtain the residuals. Instead of using daily variations $\Delta Z_n$, we used the residuals resulting from this process. These residuals represent the fluctuations of the time series once the seasonal effects are removed, thus providing a clearer insight into the underlying dynamics of the system.

Instead of subdividing the range of the residuals into fixed-width cells, we choose a different approach. We divide the range of the residuals into 500 bins, regardless of their amplitude. This fixed number of bins $z_i$ ensured a uniform distribution of data points across the entire range of the residuals. The resulting normalized histogram represents the residuals' probability density function (PDF); that is $p(z_i)$.

To calculate the $q_{\mbox{\scriptsize stat}}$ value, we plotted $\ln_q[p(z_i)]$ against $z_i^2$, where $z_i$ represents the centers of each bin in the histogram. Here, $\ln_q$ is the $q$-logarithm, defined in the same manner as in the original method. We varied the value of $q$ within an appropriate range, linearly fitting the plot for each $q$ value and evaluating the associated correlation coefficient. The optimal $q_{\mbox{\scriptsize  stat}}$ value was the one that maximized this coefficient, indicating the best linear fit for the relationship between $\ln_q[p(z_i)]$ and $z_i^2$.

Finally, with this $q_{\mbox{\scriptsize stat}}$ value, we fitted a $q$-Gaussian to the PDF of the residuals. The $q$-Gaussian was defined and adjusted in the same way as in the original method, selecting the value of $\beta$ that minimized the sum of the squares of the differences between the fitted $q$-Gaussian and the actual PDF of the residuals.

To calculate the $q_{\mbox{\scriptsize rel}}$ adapted to the deseasonalized residuals of our atmospheric series, we began by defining the temporal autocorrelation function $C(\tau)$ for the residuals, where $\tau$ is the time lag. This function is calculated using the formula
\begin{equation}
    C(\tau) = \left[\sum_{n}(z_{n+\tau} - \bar{z})(z_n - \bar{z})\right]\left[\sum_{n}(z_n - \bar{z})^2\right]^{-1},
\end{equation} where $z_n$ are the values of the residuals and $\bar{z}$ is the average of these residuals. We observed the behavior of $C(\tau)$ to determine whether it followed a $q$-exponential curve rather than a traditional exponential, defined as $q\mbox{-exp}(-x) = [1 - (1 - q) x]^{1/(1 - q)}$ for $q \neq 1$. We proceed to vary the parameter $q$ and adjust $C(\tau)$ to the $q$-exponential curve for each value of $q$, with the aim of minimizing the difference between $C(\tau)$ and the fitted $q$-exponential curve, searching for the value of $q$ that provided the best fit. The optimal value of $q$ obtained from this process, identified as the $q_{\mbox{\scriptsize rel}}$, reflects the rate at which the residuals return to a steady state after perturbations, indicating the relaxation dynamics of the system. In our study, the value of $q_{\mbox{\scriptsize rel}}$ suggests a relaxation dynamics different from that expected in classical Boltzmann-Gibbs processes.

To calculate the $q_{\mbox{\scriptsize sens}}$ adapted to the deseasonalized residuals of our atmospheric series, we followed a process similar to that described by Ferri et al (2010)\cite{Ferri2010}. Initially, we evaluated the multifractal spectrum of our time series as explained in the relevant references\cite{Ferri2010,Rutllant2003}. This approach involved determining the extremes of the multifractal spectrum, $\alpha_{\mbox{\scriptsize min}}$ and $\alpha_{\mbox{\scriptsize max}}$, which are crucial for calculating the $q_{\mbox{\scriptsize sens}}$.

In our study, we performed an extrapolation of the multifractal spectrum using a fourth-degree polynomial. This extrapolation allowed us to accurately determine the values of $\alpha_{\mbox{\scriptsize min}}$ and $\alpha_{\mbox{\scriptsize max}}$. Then, we proceeded to calculate the $q_{\mbox{\scriptsize sens}}$ using the mathematical relation given by:
\begin{equation}
\frac{1}{1 - q_{\mbox{\scriptsize sens}}} = \frac{1}{\alpha_{\mbox{\scriptsize min}}} - \frac{1}{\alpha_{\mbox{\scriptsize max}}}
\end{equation}

Applying this formula to our data, we obtained a $q_{\mbox{\scriptsize sens}}$ value indicating a sensitivity to initial conditions significantly different from what would be expected in systems that follow classical statistical mechanics. This finding emphasized the complexity and non-linear nature of the atmospheric dynamics we were analyzing and demonstrated the utility of the $q$-triplet approach in the study of complex and non-equilibrated systems.

\section{Results}

As we can observe from figure \ref{figure01}, the followings statement regarding the nature of values $q_{\text{stat}}, q_{\text{rel}}, q_{\text{sens}}$ is discussed:

\begin{itemize}

\item The $q_\text{stat}$ values range from 1.26 to 1.44. These values suggest non-Gaussian distributions in the atmospheric data. The variations in $q_\text{stat}$ might indicate different degrees of complexity or long-range correlations in the atmospheric dynamics at different times of the day.

\item The $q_\text{rel}$ values range from 1.91 to 4.61, represent the rate of relaxation of the system towards equilibrium. Higher values, especially those significantly greater than 1, imply a slower relaxation process, suggesting that the atmospheric system might be far from equilibrium and may exhibit complex relaxation dynamics.

\item The $q_\text{sens}$ values range from -0.23 to 0.32. This parameter measures the sensitivity of the system to initial conditions, indicating how small changes can influence the system's future state. Negative values imply reduced sensitivity to initial conditions, whereas positive values indicate increased sensitivity.

\item The range of values for all three parameters of the $q$-triplet method suggests that Antofagasta's atmospheric dynamics are complex and non-linear. This complexity is evident from the deviations from classical statistical mechanics and indicates that standard models might not fully capture the behavior of this atmospheric system (see Table 1).
\end{itemize}

\begin{figure}[h!]
\begin{center}
\includegraphics[width=18cm,trim={5.5cm 8.4cm 8cm 9.3cm},clip]{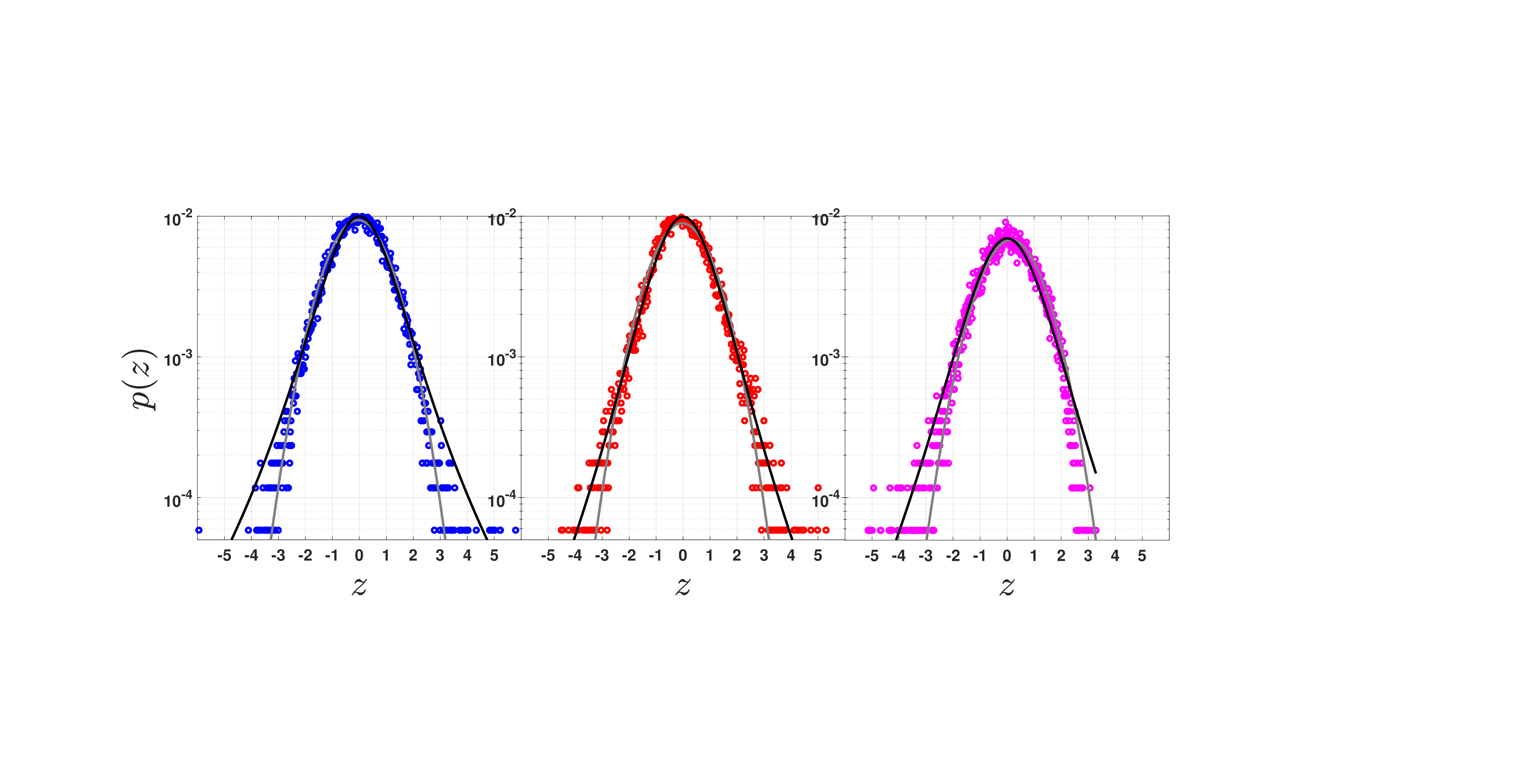}
\end{center}
\end{figure}

\begin{figure}[h!]
\begin{center}
\includegraphics[width=18cm,trim={5.5cm 8.4cm 8cm 9.3cm},clip]{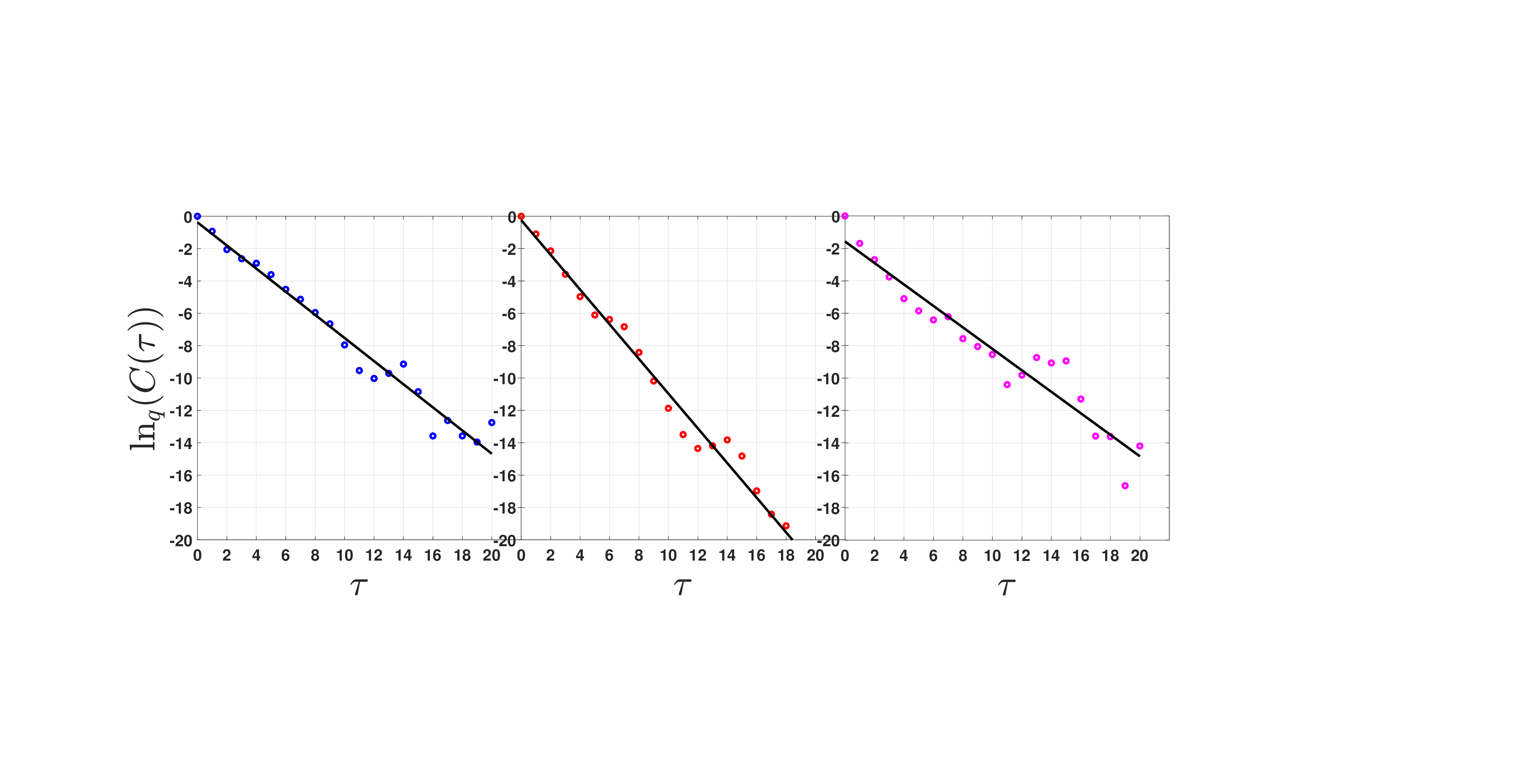}
\end{center}
\end{figure}

\begin{figure}[h!]
\begin{center}
\includegraphics[width=18cm,trim={5.5cm 8.4cm 8cm 9.3cm},clip]{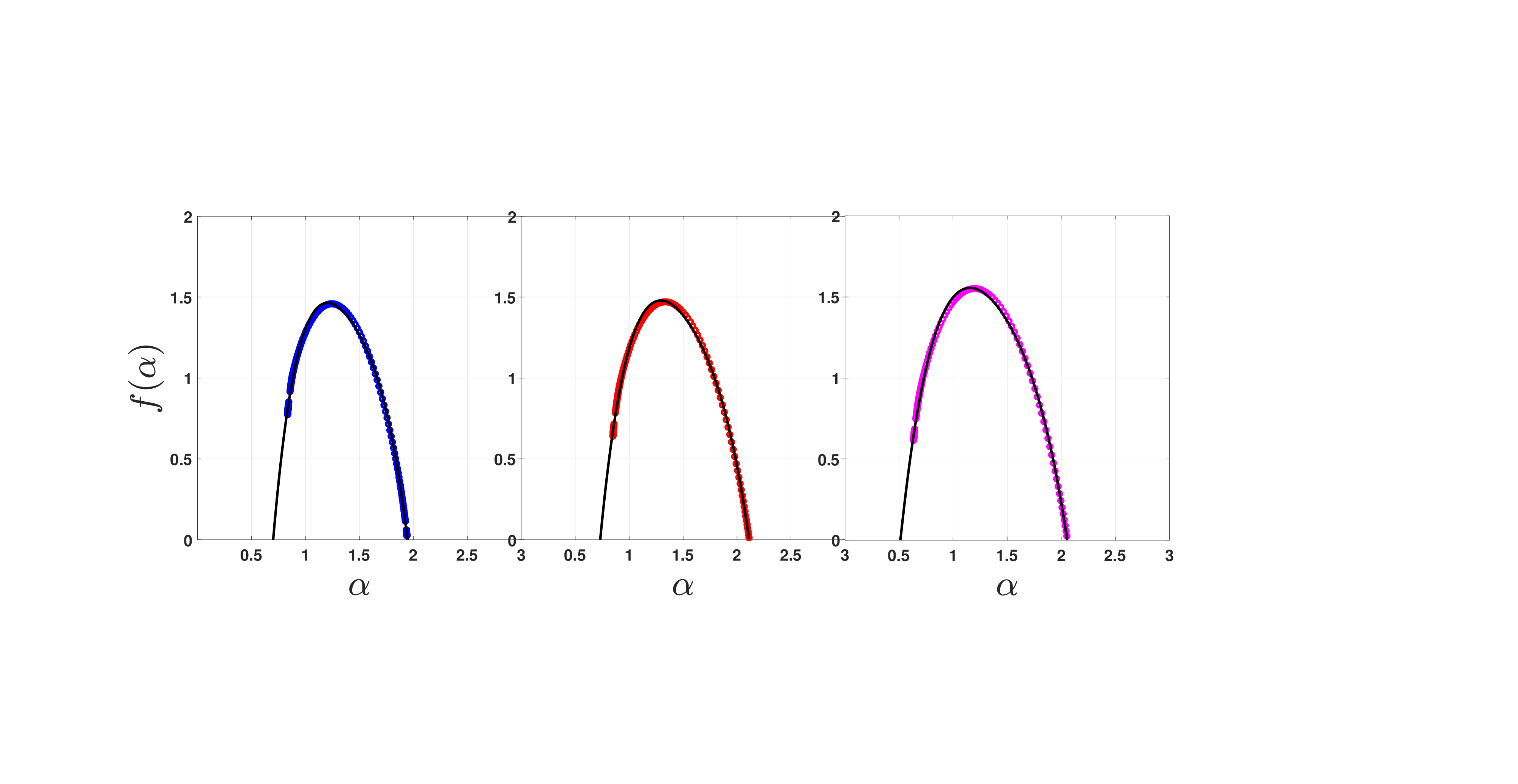}
\end{center}
\caption{\label{figure01} Illustration of $q$-triplet calculation at 08:00. Top row: Probability density functions $p(z)$ with q-Gaussian fits for atm. pressure (blue), air temperature (red), and rel. humidity (magenta). Middle row: (Log) Autocorrelation function $\ln_q[C(\tau)]$ versus time lag $\tau$, showing linear fits indicating relaxation dynamics. Bottom row: Multifractal spectra $f(\alpha)$ demonstrating multifractality of data.}
\end{figure}

\begin{table}[h!]
\centering
\begin{tabular}{lccc}
\hline
\textbf{Variable} & $q_{\mbox{\scriptsize stat}}$ & $q_{\mbox{\scriptsize rel}}$ & $q_{\mbox{\scriptsize sens}}$ \\
\hline
TT08 & 1.30 & 4.61 & -0.12 \\
TT14 & 1.44 & 4.60 & -0.16 \\
TT20 & 1.28 & 4.15 & -0.23 \\
PA08 & 1.39 & 2.41 & -0.10 \\
PA14 & 1.28 & 2.42 & \ 0.14 \\
PA20 & 1.26 & 2.35 & \ 0.14 \\
HR08 & 1.30 & 1.91 & \ 0.32 \\
HR14 & 1.29 & 2.89 & \ 0.15 \\
HR20 & 1.29 & 2.19 & \ 0.28 \\
\hline
\end{tabular}
\caption{Compilation of $q$-triplet values for atmospheric pressure (PA), air temperature (TT), and relative humidity (HR) measured at three different times: 08:00, 14:00, and 20:00.}
\label{table:your_label_here}
\end{table}

\vspace{4mm}

\section{Concluding remarks}
In this work, we utilized the q-triplet method to conduct a comprehensive analysis of the atmospheric time series data from Antofagasta, overcoming the challenges posed by missing data. Our analysis uncovered pronounced non-linearity and complexity in the atmospheric dynamics of the region, deviating from traditional statistical models, particularly ARIMA models. Notably, the data demonstrated a q-Gaussian distribution pattern and a q-exponential decay in autocorrelation, indicating a substantial long-term dependence. These findings underscore the limitations of standard linear models and highlight the necessity for more sophisticated techniques capable of addressing the inherent non-stationarity and the complex error structures typical in such systems. Moving forward, we plan to expand our methodology to include additional regions and diverse climatic variables, and to develop innovative approaches to overcome the prevalent data limitations in climate research. This work significantly enhances our understanding of Antofagasta's climate and sets the stage for future investigations into complex dynamic systems.

\ack
We would like to thank Agencia Nacional de Investigación y Desarrollo (ANID), Chile (FONDECYT 1211848; ANILLO ATE220022); Pontificia Universidad Católica de Valparaíso (PUCV), Chile (Project 123.774/2021), “Núcleo de Investigación No.7 UCN-VRIDT 076/2020, Núcleo de modelación y simulación científica (NMSC)” and ``Núcleo de Investigación No.2 UCN-VRIDT 042/2020, Sistemas Complejos en Ciencia e Ingeniería".

\end{document}